\documentclass[fleqn,usenatbib]{mnras}

\usepackage{newtxtext,newtxmath}

\usepackage[T1]{fontenc}
\usepackage{ae,aecompl}

\usepackage{graphicx}	
\usepackage{amsmath}	
\usepackage{amssymb}	
\usepackage{subfig}
\usepackage{lipsum}
\usepackage{pdflscape}
\usepackage{booktabs}
\usepackage{multirow}
\usepackage{float}
\usepackage{xr}

\newcommand{\msun}{M_{\odot}}
\newcommand{\rsun}{R_{\odot}}
\newcommand{\mbh}{M_{\rm BH}}

\newcommand{\msgr}{M_{\rm Sgr}}

\newcommand{\mc}{M_{\rm c}}

\title[Detecting Intermediate-mass Black Holes]{Astrometric Detection of Intermediate-Mass Black Holes At the Galactic Centre}

\author[E. Girma and A. Loeb]{
Eden Girma,$^{1}$\thanks{E-mail: edengirma@live.com}
and Abraham Loeb$^{1}$
\\
$^{1}$Department of Astronomy, Harvard University, 60 Garden St., Cambridge, MA 02138
}

\date{Accepted 2018 September 21. Received 2018 September 14; in original form 2018 July 6}

\pubyear{2018}

\begin{document}
\label{firstpage}
\pagerange{\pageref{firstpage}--\pageref{lastpage}}
\maketitle

\begin{abstract}
We assess the astrometric detectability of intermediate-mass black holes populating the inner parsec of the Milky Way Galaxy. The presence of these objects induces dynamical effects on Sgr A* and the star S2, which could be detected by next generation astrometric instruments that enable micro-arcsecond astrometry. An allowed population of ten $10^4~\msun$ IMBHs within one parsec induces an angular shift of $\sim 65~\mu\text{as yr}^{-1}$ on the position of Sgr A*, corresponding to a perpendicular velocity component magnitude of $1.6~\text{km s}^{-1}$.  It also induces changes in the orbit of S2 that surpass those induced by general relativity but lie within observational constraints, generating a mean angular shift in periapse and apoapse of 62 $\mu$as and 970 $\mu$as respectively.

\end{abstract}




\section{Introduction}
\label{sec:1}

The mass distribution of detected black holes exhibits an absence of \textit{intermediate-mass black holes} (IMBHs), with masses of $10^2-10^5~\msun$. These black holes are too massive to form from the collapse of present-day stars, their dynamical effect are too weak to be easily detectable, and their supposed environments lack the high density and velocities that could potentially fuel growth. Despite the relatively weak observational evidence available, IMBHs continue to represent a critical missing link between stellar and supermassive black holes, and a definitive discovery could provide essential insights into black hole formation, evolution, and dynamics.

There is a range of models for IMBH formation, including the direct collapse of very massive Population III stars \citep{Bromm:1999,Abel:2000,Figer:1996}, gas accretion onto stellar-mass black holes \citep{Leigh:2013}, and stochastic formation via black hole mergers and mass transfers in black hole binaries \citet{Giersz:2016}. Dense stellar clusters are also proposed to form IMBHs through the collapse of massive runaway mergers \citep{Ebisuzaki:2001, Zwart:2002}; though such mergers are likely to experience severe mass loss though stellar winds, collapsing to black holes of mass $\lesssim 250~\msun$ \citep{Petts:2017}. These dense clusters are characterised by a relaxation time, $\tau_r$, during which a test object changes its velocity by an order itself due to random perturbations induced by surrounding gravitational fields \citep{Chandrasekhar:1941},
\begin{equation}
	\tau_r \sim \frac{\sigma^3}{8\pi G^2 \mu^2 n \ln \Lambda}\label{relaxation}
\end{equation}
where $\mu$ is stellar mass, $\sigma$ is the velocity dispersion, $n$ is the number of stars per unit volume, and $\Lambda$ is the ratio of the maximal and minimal impact parameters, $b_{\rm max}$ and $b_{\rm min}$.  For a virialised and self-gravitating system with average stellar mass $\langle m_* \rangle$, the typical $b_{\rm min}$ (which leads to a deflection angle of $\pi / 2$) is $G\langle m_* \rangle / \sigma^2$, and $b_{\rm max}$ is on the order of the half-mass radius $R_h$ \citep{Freitag:2006}. A larger mass object $M$, travelling with velocity $v_M$ and embedded in spherically symmetric distribution of lower-mass objects $m$ with velocities $v_m < v_M$, will experience a decelerating force characterised by the \citeauthor{Chandrasekhar:1943} dynamical friction formula
\begin{equation} \label{dynamicalfriction}
	\frac{d\vec{v}_M}{dt} = -16\pi^2 \ln\Lambda G^2m(M+m)\frac{\int_0^{v_M}f(r,v)v^2 dv}{v_M^3}\vec{v}_m
\end{equation}
inducing migration of this larger mass to the centre of the system over time. In this vein, globular clusters of mass $M$ orbiting a host galaxy centre with circular velocity $v_c$ eventually spiral inwards from an initial radius $r_i$ after a time \citep[][eq. 7-25]{Binney:2011}
\begin{align}\label{clusterfric}
t_{fric} &= \frac{1.17}{\ln \Lambda}~\frac{r_i^2 v_c}{GM} = 2.64 \times 10^{10}~\text{yr}~\left( \frac{\ln \Lambda}{10}\right)^{-1} \left( \frac{r_i}{2~\text{kpc}} \right)^2 \times \\
&\; \times \left( \frac{v_c}{250~\text{km s}^{-1}} \right) \left( \frac{M}{10^6~\msun} \right)^{-1}, \nonumber
\end{align}
theoretically populating the centre of the Milky Way with stars, stellar remnants, and IMBHs.

Supposing the existence of a surrounding IMBH population in the Galactic Centre, dynamical friction is believed to segregate these higher mass objects into a density cusp proportional to $r^{-7/4}$ \citep{Morris:1993}.  These processes of relaxation and mass segregation have not only been demonstrated by various $N$-body codes simulating the dynamics of IMBHs in dense or in-spiralling stellar clusters \citep[see][]{Zwart:2002, Gurkan:2004, Antonini:2012, Mastrobuono:2014}, but may also provide a necessary explanation for the perplexing number of massive main-sequence stars present around the centre of the Milky Way \citep[the so-called ``paradox of youth''; see][]{Ghez:2003, Paumard:2006, Ghez:2008, Lu:2008, Yelda:2014}. The orbital properties of these young stars reflect what would result from millions of years' interaction with an IMBH of mass $\gtrsim 1500~\msun$, further supporting an IMBH transfer model \citep{Merritt:2009, Fujii:2009, Fujii:2010}. Additional models for young stellar populations in the Galactic Centre suggest in-situ formation \citep{Bonnell:2008, Mapelli:2008, Alig:2011, Alig:2013} or tidal disruption of stellar binaries \citep{Miller:2005, Bromley:2006, Perets:2007, Perets:2009}.

The central supermassive black hole of the Milky Way (Sgr A*) will experience reflex motion due to gravitational perturbations, which may be distinguishable in magnitude from changes induced by surrounding stars \citep{Merritt:2007,Chatterjee:2002,Loeb:2013, Hansen:2003, Yu:2003}. The star S2, which orbits Sgr A* with a semi-major axis of $\sim 5$ milliparsecs (mpc) \citep{Gillessen:2009}, might also hint at the existence of IMBHs through changes in eccentricity and orbital plane that surpass effects of general relativity \citep{Gualandris:2010,Weinberg:2005}. Hence, high-accuracy measurements of residual proper motion of Sgr A* so far quantified by \citet{Reid:2004} and evolving orbital parameters of S2 are essential in helping discern whether the inner parsec hosts IMBHs.

In what follows, we investigate whether particular IMBH populations in the Galactic Centre, characterised by various density profiles and black hole masses, produce a gravitational effect on Sgr A* and S2 that is both detectable astrometrically and differentiable from the effects of surrounding stars. Using model density profiles for stellar- and intermediate-mass black holes (Section \ref{sec:3}), we developed a code that used the open-source $N$-body integrator \texttt{REBOUND} \citep{Rein:2012} to create a simulated orbital system involving Sgr A*, a surrounding population of black holes, and the star S2 (Section \ref{sec:2}). S2 is analysed over other S-stars, as it is one of the brightest of its counterparts and possesses an orbital period within 16 years. With this, its observed perturbations are particularly useful in characterising the surrounding gravitational potential. Note that the effects of IMBH perturbations are necessarily expected on other S-stars, and our code is made adaptable such that one could model the effects of an IMBH population on an S-star of one's choosing.

By simulating scenarios in which the central parsec is populated or not populated with black holes, we identify in Section \ref{sec:4} the dynamical signature of the IMBHs.  Section \ref{sec:5} summarises our findings in the context of current observational constraints, and discusses how they may inform future observations by the Event Horizon Telescope (EHT)\footnote{https://eventhorizontelescope.org/} and GRAVITY\footnote{http://www.mpe.mpg.de/ir/gravity/} instrument on the VLT data collections on the angular displacements of Sgr A* and the orbit of S2.

\section{Numerical Methods}
\label{sec:2}

\begin{table*}
\centering
\begin{tabular}{@{}lccccc@{}}
\toprule
& $\langle m_{\rho} \rangle=1~\msun$ & $\langle m_{\rho} \rangle=10~\msun$ & $\langle m_{\rho} \rangle=100~\msun$ & $\langle m_{\rho} \rangle=10^3~\msun$ & $\langle m_{\rho} \rangle=10^4~\msun$ \\ \midrule
\underline{Error threshold} \\
$\;\;\;$ Discrete & $10^{-4}$ & $10^{-7}$ & $10^{-7}$ & $10^{-7}$ & $10^{-7}$\\
$\;\;\;$ Smooth & $10^{-4}$ & $10^{-5}$ & $10^{-5}$ & $10^{-4}$ & $5 \times 10^{-3}$\\
\underline{$N_{\rm runs}$ successful} & 309 & 413 & 1299 & 1394 & 943\\
\underline{$N_{\rm particles}$}\\
$\;\;\;$ Mean & 31922 & 11462 & 358 & 4 & 10\\
$\;\;\;$ Max & 32069 [0.3\%] & 11623 [0.2\%] & 405 [0.1\%] & 5 [17.8\%] & 14 [1.5\%]\\
$\;\;\;$ Min & 31770 [0.3\%] & 11272 [0.2\%] & 321 [0.1\%] & 0 [0.1\%] & 5 [0.5\%] \\ \bottomrule
\end{tabular}
\caption[Information on various simulation parameters]{The error threshold determining which simulations of discrete and smooth density profiles were analysed, number of successful simulation runs (out of 1500), and the average, minimum, and maximum number of particles drawn from the tested density profiles. Bracketed percentages indicate the percentage of runs that involved the maximum or minimum number of particles.}\label{tab:sim-data}
\end{table*}

To simulate the various $N$-body models involving Sgr A*, S2, and a surrounding population, we developed a \texttt{Python} package using the open-source, multi-purpose $N$-body software package \texttt{REBOUND} \citep{Rein:2012}.  \texttt{REBOUND} is designed to integrate a variety of gravitational systems and supporting both collisional and collision-less (classical) dynamics; upon importing the package, simulations may be initialised and performed via a provisional \texttt{Python} module.  The \texttt{Python} package created for the purposes of this work used \texttt{REBOUND} to randomly realise and integrate two types of orbital systems: (i) Sgr A* and S2 along with varied IMBH and stellar populations, and (ii) Sgr A* and S2 affected by a \textit{smooth} density profiles matching those of the examined surrounding populations. These systems were differentiated to properly investigate how the effects of discrete objects on the orbital parameters of S2 differ from that of a smoothly distributed mass.

Each simulation is initialised with two active particles: a central black hole of mass $M_{\rm cbh} = 4.4 \times 10^6~\msun$ resting at the origin, and a star of mass $20~\msun$ representing S2, with orbital parameters provided by \citet{Gillessen:2009}. To add particles of mass $m$ reflecting a density profile $\rho$ that extends within a narrow range from a distance $r_{\rm min}$ to $r_{\rm max}$, we first calculate the number of particles present in the simulation $N_{\rm p}$. This is obtained by drawing randomly from a Poisson distribution with mean
\begin{equation}
	N_{\rm mean} = \frac{1}{m}\left(M(r_{\rm max})-M(r_{\rm min})\right),
\end{equation}
where 
\begin{equation}
M(r) = \int_0^r 4\pi s^2 \rho(s) ds
\end{equation} describes the total mass contained within a radius $r$ for a density profile $\rho$. To distribute these particles according to their density profile, we randomly determine the distance of each particle from the origin through an inverse transform sampling method. The appropriate cumulative density function for particle position is \[f(r) = M(r)/(\mc)\] where $\mc$ is the total mass contained in the density profile $\rho$. We uniformly draw a value $u$ from the interval $(f(r_{\rm min}),f(r_{\rm max}))\subset [0,1]$, let $m_u = \mc * u$, and calculate the particle position as $r = f^{-1}(m_u)$. The unit directional vector $\hat{\mathbf{r}}=(x,y,z)$ of the particle position is then determined via sphere point picking: given random variate $u \in [-1,1]$ and $\theta \in [0,2\pi)$,
\begin{align}
	x &= \sqrt{1 - u^2} \cos \theta,\\
	y &= \sqrt{1 - u^2} \sin \theta,\\
	z &= u.
\end{align}

The phase space distribution of the orbital system is used to randomly determine the velocity of each particle at a known position $\mathbf{r}$. This distribution function is approximated as the analytic solution found by \citet{Tremaine:1994} that corresponds to a Hernquist profile with a central black hole:
\begin{equation} \label{eq:herndf}
f(\mathcal{E}) = \frac{2\Gamma(2) }{2^{7/2} \pi^{5/2} (G\mc a)^{3/2} \mu \Gamma(\frac{1}{2})} \mathcal{E}^{-1/2}.
\end{equation}
where $\mc$ is the total cluster mass, $\mu$ is the ratio of central black hole mass to cluster mass, and $\Gamma$ the gamma function. Within a sufficiently thin spherical shell the phase space distribution depends only on velocity.  Hence from Equation \eqref{eq:herndf}, we inversely derived continuous density functions describing the distribution of bound velocities ($v < \sqrt{-2\Phi}$) for $10^3$ spherical shells of thickness $10^{-3}~r_{\rm max}$ situated between $r_{\rm min}$ and $r_{\rm max}$.  The speed of a particle at position $\mathbf{r}$ is randomly drawn from the continuous density function in the corresponding spherical shell. Upon drawing this magnitude, the direction of the velocity vector is randomly determined and ensured to be perpendicular to the position vector.  The particles are then added to the simulation, specified by the orbital elements calculated from their position and velocity vectors, such that they are gravitationally interacting with the central massive object and the S-star, but not with each other.

For simulations testing the effects of smooth density profiles, we do not add any other particles besides Sgr A* and S2. Instead, we include an additional force induced by the smooth density profile in question.

For each density profile, we performed 1500 different random realisations of the associated particle system surrounding Sgr A* and S2.  $N$-body simulations were advanced using the hybrid symplectic integrator \texttt{Mercurius}, which combines the symplectic \texttt{WHFast} and implicit \texttt{IAS15} integrator modules already built into \texttt{REBOUND} \citep{Rein:2014,Rein:2015}.  The properties of the integrator were further set such that the simulation merges particles that collide with each other, preserving mass and momentum, and tracks energy that is lost due to collisions.  Error is kept at a minimum by identifying the optimal time step for each simulated scenario, which ranged from 20 hours to 80 hours. Between each call of the integration method, we record the position and velocity Cartesian components of Sgr A* and S2, the various orbital elements of S2 ($a$, $e$, $i$, $\Omega$, and $\omega$), and the total number of particles remaining in the simulation. Relative energy error was calculated, and simulations reaching energy errors greater than $10^{-7}$ ($1 \times 10^{-4}$ for stellar control) were discarded from the analysed data-set. The number of successful runs, along with the average, maximum, and minimum number of particles for each simulated model are additionally summarised in Table \ref{tab:sim-data}.

\section{Theoretical Models}
\label{sec:3}

The numerical simulations described in Section \ref{sec:2} examined five distinct density profiles: a stellar control, stellar-mass black holes of $10~\msun$ and $100~\msun$, and intermediate-mass black holes of $10^3~\msun$ and $10^4~\msun$.  Note that although we choose to include the IMBH mass of $10^4~\msun$ in this work, IMBHs of mass $\gtrsim 10^4~\msun$ have been largely excluded from most parameters in the Galactic Centre \citep[see][]{Reid:2004, Gualandris:2009}.

The stellar control is described by a \citet{Hernquist:1990} spherical density profile
\begin{equation}
\rho_{\rm H}(r) = \frac{\mc}{2\pi} \frac{a}{r(r+a)^3}
\label{hernprofile}
\end{equation}
where $\mc = 4.4 \times 10^6~\msun$ is the total cluster mass, and $a = 1~\text{pc}$ the scale length. We adopt this profile as a relatively close approximation to the actual broken law stellar density in the inner parsec, provided by \citet{Schodel:2007} as
\begin{equation} \label{stellardensity}
	\rho_*(r) \simeq (1.7 \pm 0.8) \times 10^6 \left(\frac{r}{0.22~\text{pc}}\right)^{-\gamma}~\msun\text{pc}^{-3},
\end{equation}
where $\gamma = 1.2$ for $r < 0.22~$pc and $\gamma = 1.75$ for $r > 0.22~$pc. We assume a stellar characteristic mass and radius of $1~\msun$ and $1~\rsun$ respectively.  The inner cut off radius for this profile is set as the tidal radius, \begin{align*}\label{tidalradius}
r_t &\simeq R_{\star}(M_{\rm BH}/M_{\star})^{1/3}\\
&= 0.76~\text{AU} \left(\frac{R_{\star}}{\rsun}\right) \left(\frac{M_{\rm BH}}{4.4\times 10^6~\msun}\right)^{1/3} \left(\frac{M_{\star}}{\msun}\right)^{-1/3}.
\end{align*}
The displacement of Sgr A* over year-long timescales is dominated by components orbiting around 0.01 pc, with analytic treatments finding proper motions of $\sim 0.07~$km s$^{-1}$ when considering stars within 2 pc of the Galactic Centre. \citep{Broderick:2011, Reid:2004}. Thus, to shorten computation time while maintaining reasonable accuracy on our results, we simulate the stellar density profile out to 0.1 pc.

Density profiles for stellar-mass black holes are adapted from the methods of \citet{Miralda:2000}, which construct a density profile of stellar remnants that is necessarily a fraction of that of the surrounding stellar cluster: for a characteristic black hole mass $\mbh$ and stellar density profile $\rho_*(r)$, \begin{equation}
  \rho_{\rm SBH}(r) = C\rho_*(r),\; C \equiv \left( \frac{0.23~\msun}{\langle \mbh \rangle} \right)^{\frac{1}{2}},
\end{equation}\label{SBH-density-profile}
where $\mbh$ is the black hole mass.  We set $\rho_*(r) = \rho_{\rm H}(r)$ from Equation \eqref{hernprofile} and $\mbh = 10~\msun \text{ or } 100~\msun$. 

In identifying a potential density profile IMBHs of mass $\sim 10^3~\msun$, we consider the results of $N$-body simulations performed by \citet{Zwart:2006} that predict $\sim 50$ such black holes will be present within $10~$pc of Sgr A*.  Given that general estimations of the final mass-segregated distribution of stellar remnants range from flat-cores to extreme cusps \citep[e.g]{Alexander:2009,Merritt:2010}, we adopt a fiducial $\rho \propto r^{-2}$ model:
\begin{equation}
  \rho_{\rm Z06}(r) = 3.98 \times 10^2 \left(\frac{r}{\text{pc}} \right)^{-2}~\msun\text{pc}^{-3}
\end{equation}\label{Z06-density-profile}

For IMBHs of mass $\sim 10^4~\msun$, we use the density profile obtained by \citet{Mastrobuono:2014} in their work simulating twelve in-spiralling nuclear star clusters:
\begin{equation}
  \rho_{\rm M14}(r) = 6.20\times 10^3\left(\frac{r}{\text{pc}} \right)^{-2.32}~\msun\text{pc}^{-3}.
\end{equation}\label{M14-density-profile}
Given that the above density profiles diverge as $r$ approaches 0, an inner cut-off radius is defined as the semi-major axis $a_{\rm c}$ associated with a gravitational wave time for coalescence $T_c$ equalling Hubble time \citep{Peters:1964}, 
\begin{align} \label{GW-scale}
a &\geq 4\left( \frac{1}{5} \frac{G^3 (M_{\rm Sgr}+M_{\rm BH}) M_{\rm Sgr} M_{\rm BH}}{c^5}T_{\rm c}\right)^{1/4}\\
&= 0.7~\text{mpc} \left( \frac{M_{\rm BH}}{10^3~\msun} \right)^{1/4}. \nonumber
\end{align}

\begin{table*}
\centering
\begin{tabular}{@{}lcccccccccc@{}}
\multicolumn{11}{c}{\textbf{Angular displacement of Sgr A*} ($\mu$as)}\\*[2pt]
\toprule
& \multicolumn{2}{c}{$\langle m_{\rho} \rangle=1~\msun$} & \multicolumn{2}{c}{$\langle m_{\rho} \rangle=10~\msun$} & \multicolumn{2}{c}{$\langle m_{\rho} \rangle=100~\msun$} & \multicolumn{2}{c}{$\langle m_{\rho} \rangle=10^3~\msun$} & \multicolumn{2}{c}{$\langle m_{\rho} \rangle=10^4~\msun$} \\ \cmidrule(l){2-3} \cmidrule(l){4-5} \cmidrule(l){6-7} \cmidrule(l){8-9} \cmidrule(l){10-11}
& $\mu$ & $\sigma$ & $\mu$ & $\sigma$ & $\mu$ & $\sigma$ & $\mu$ & $\sigma$ & $\mu$ & $\sigma$ \\ \midrule

$\;\;\;1~$yr & 0.7 & 0.3 & 1.2 & 0.6 & 2.4 & 1 & 2.8 & 1.9 & 67.8 & 33.8\\
$\;\;\;5~$yr & 5.8 & 1.66 & 6.9 & 2.8 & 11.9 & 5 & 13.6 & 9.2 & 333.8 & 165.8\\
$\;\;\;10~$yr & 14.4 & 3.4 & 12.7 & 5.6 & 23.6 & 10 & 26.8 & 17.8 & 654.9 & 321.2\\
$\;\;\;15~$yr & 22.7 & 5 & 20.6 & 8.5 & 35.5 & 14.9 & 39.6 & 25.7 & 953.9 & 459\\ \bottomrule
\end{tabular}
\caption{Mean ($\mu$) and standard deviation ($\sigma$) of the angular displacement of Sgr A*, measured relative to its original position, induced by discrete density profiles describing the distribution of stars ($\langle m_{\rho} \rangle=1~\msun$), stellar black holes ($\langle m_{\rho} \rangle=10~\msun,\;100~\msun$), and intermediate mass black holes ($\langle m_{\rho} \rangle=1000~\msun,\;10^4~\msun$). See  Figures \ref{fig:1hist}, \ref{fig:10hist}, \ref{fig:100hist}, \ref{fig:1000hist}, and \ref{fig:10000hist} for its visual representation.}
\label{tab:Sgr-displacement}
\end{table*}

\begin{table*}
\centering
\begin{tabular}{@{}lcccccccccc@{}}
\toprule
& \multicolumn{2}{c}{$\langle m_{\rho} \rangle=1~\msun$} & \multicolumn{2}{c}{$\langle m_{\rho} \rangle=10~\msun$} & \multicolumn{2}{c}{$\langle m_{\rho} \rangle=100~\msun$} & \multicolumn{2}{c}{$\langle m_{\rho} \rangle=10^3~\msun$} & \multicolumn{2}{c}{$\langle m_{\rho} \rangle=10^4~\msun$} \\ \cmidrule(l){2-3} \cmidrule(l){4-5} \cmidrule(l){6-7} \cmidrule(l){8-9} \cmidrule(l){10-11}
& $\mu$ & $\sigma$ & $\mu$ & $\sigma$ & $\mu$ & $\sigma$ & $\mu$ & $\sigma$ & $\mu$ & $\sigma$ \\ \midrule
$|v|$ (km s$^{-1}$) & 0.06 & 0.02 & 0.05 & 0.02 & 0.09 & 0.04 & 0.1 & 0.06 & 2.63 & 1.33\\
$|v_{\perp}|$ (km s$^{-1}$) & 0.04 & 0.01 & 0.03 & 0.01 & 0.05 & 0.02 & 0.06 & 0.04 & 1.57 & 0.86\\ \bottomrule
\end{tabular}
\caption[Mean and standard deviation of absolute magnitude and perpendicular component magnitude of intrinsic velocity induced on Sgr A*]{Mean ($\mu$) and standard deviation ($\sigma$) of the absolute intrinsic velocity and its perpendicular component of Sgr A*, induced by discrete density profiles describing the distribution of stars ($\langle m_{\rho} \rangle=1~\msun$), stellar black holes ($\langle m_{\rho} \rangle=10~\msun,\;100~\msun$), and intermediate mass black holes ($\langle m_{\rho} \rangle=1000~\msun,\;10^4~\msun$). The table is visualized in  Figures \ref{fig:1vel}, \ref{fig:10vel}, \ref{fig:100vel}, \ref{fig:1000vel}, and \ref{fig:10000vel}.}
\label{tab:Sgr-velocity}
\end{table*}
\section{Simulation Results}
\label{sec:4}

\subsection{Measured Gravitational Effect on Sgr A*}

We calculated the relative angular deviation of Sgr A* measured at simulation times $t = 1~\rm yr$, $5~\rm yr$, $10~\rm yr$, and $15~\rm yr$, along with the velocity magnitude and velocity component magnitude perpendicular to the Galactic plane.  To compute this perpendicular component, we assume it is equal on average to the additional orthogonal velocity components, implying $v_{\perp} = v/\sqrt{3}$ where $v$ is the total 3D velocity amplitude. The result summaries are further supplemented by Tables \ref{tab:Sgr-displacement} and \ref{tab:Sgr-velocity}, which list the mean and standard deviation of the distributed values. In converting distance to subtended angle, we assume a distance from the galactic centre of $R_0 = 8.0~$kpc \citep{Reid:2004}.

For the run with background stars, the distribution of the angular shift of Sgr A* relative to its initial position after a time period of 15 years was best fit by a gamma distribution with mean $22.7 \pm 5.0~\mu$as. The distribution of velocity magnitudes for Sgr A* due to the presence of this profile was fit by a Maxwell-Boltzmann distribution with mean $0.06 \pm 0.02~\text{km s}^{-1}$, and a perpendicular component magnitude of $0.04 \pm 0.01~\text{km s}^{-1}$.

\begin{figure}[h]
	\centering
    \includegraphics[width=0.45\textwidth]{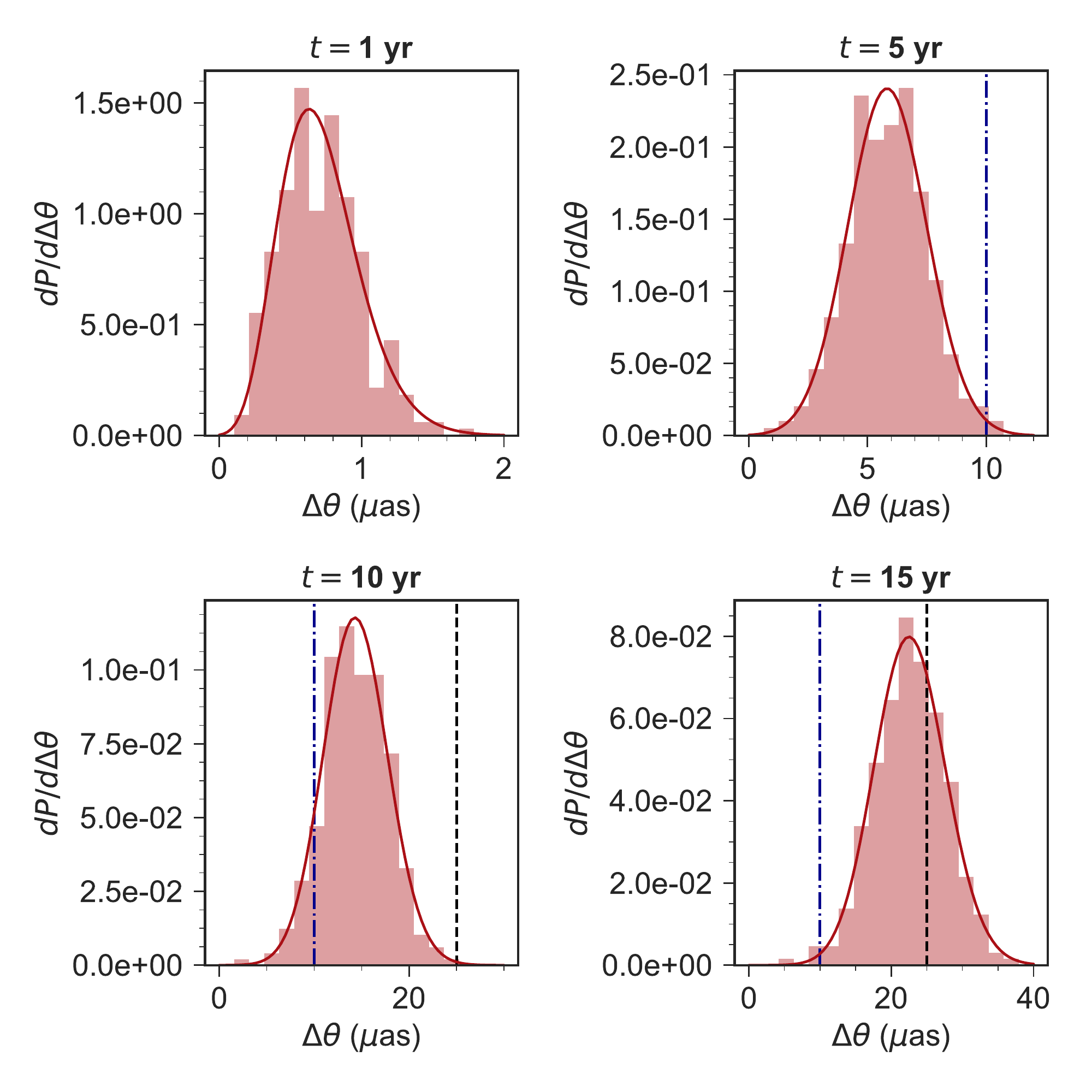}
    \caption{Probability distribution of angular displacement ($\Delta \theta$) of Sgr A* at four times ($\Delta t=1~$yr, $5~$yr, $10~$yr, and $15~$yr), when surrounded by stellar control. Each histogram is normalized to unity and fitted to a gamma distribution (marked by the solid curve) with mean and standard deviation recorded in Table \ref{tab:Sgr-displacement}.  Blue dot-dashed lines and black dashed in the histogram of angular displacement indicate the potential angular precision of GRAVITY ($10~\mu$as) and the EHT ($25~\mu$as) respectively.}
    \label{fig:1hist}
\end{figure}

\begin{figure}[h]
	\centering
    \includegraphics[width=0.45\textwidth]{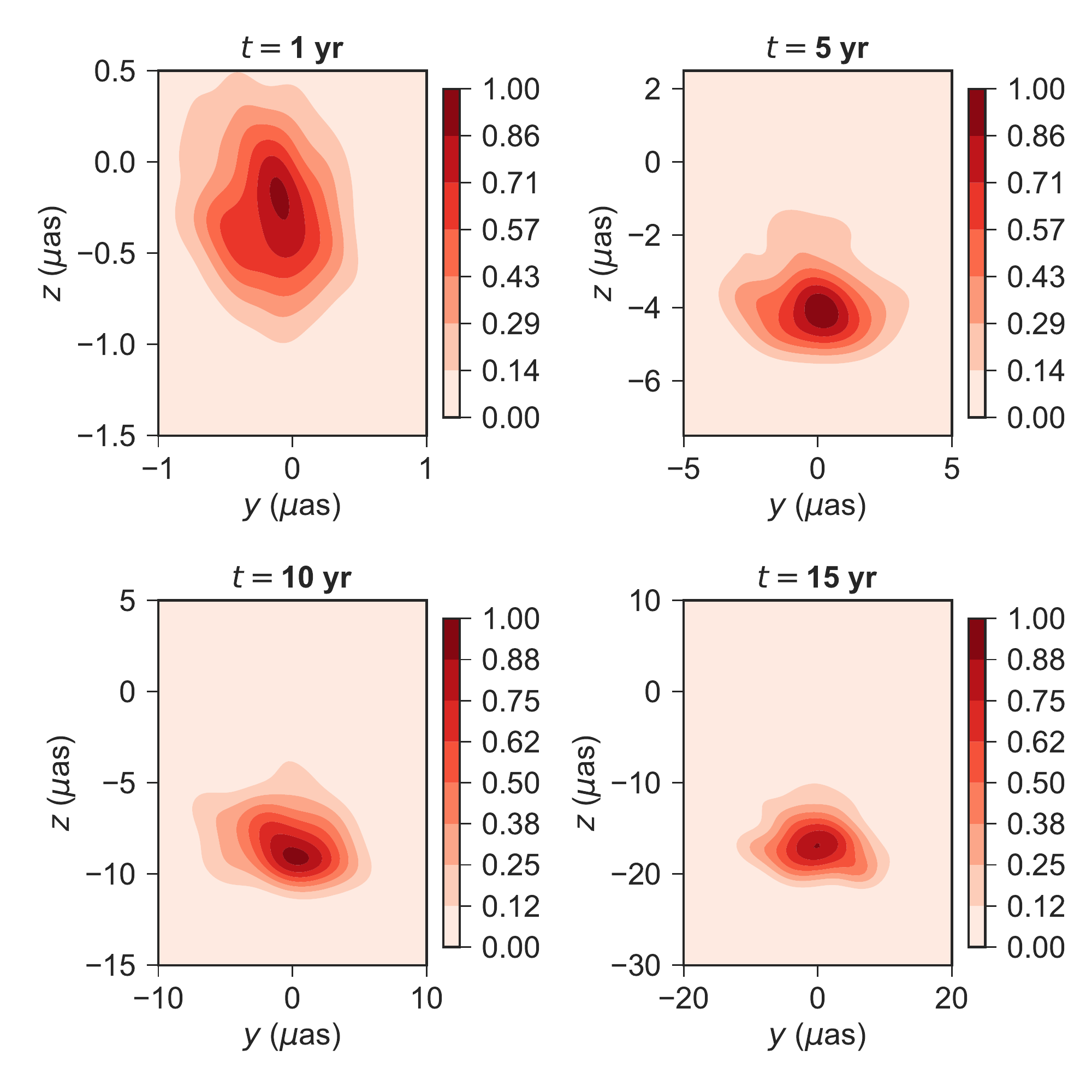}
    \caption{Bivariate probability density of Sgr A* at four times ($\Delta t=1~$yr, $5~$yr, $10~$yr, and $15~$yr), when surrounded by stellar control.}
    \label{fig:1kde}
\end{figure}

\begin{figure}[h]
	\centering
    \includegraphics[width=0.45\textwidth]{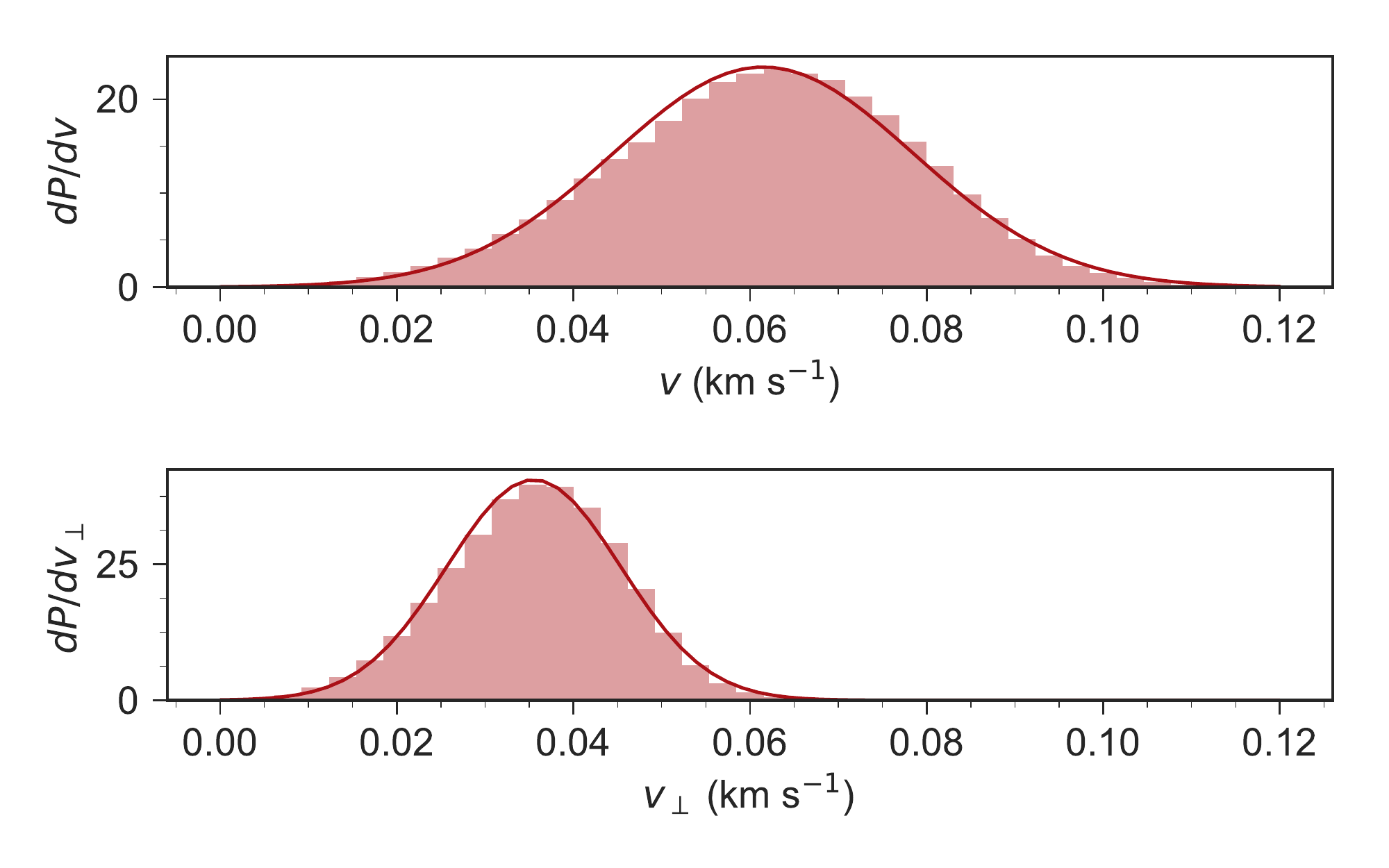}
    \caption{Probability distribution of velocity of Sgr A* when surrounded by stellar control. The upper panel displays the distribution of total velocity magnitude ($v$) measured at each time step of a simulation, while the lower panel shows the distribution of measured velocity component perpendicular to the galactic disk ($v_{\perp}$). Solid lines indicate fitted distributions with means and standards deviation noted in Table \ref{tab:Sgr-velocity}. Total velocity magnitude is fitted to a Maxwell-Boltzmann distribution while perpendicular velocity magnitude is fitted to a gamma distribution, and the area under each curve is normalized to unity.}
    \label{fig:1vel}
\end{figure}

An average of 11462 $10~\msun$ black holes and 358 $100~\msun$ black holes were drawn within the inner parsec, displacing Sgr A* on average by $20.6\pm 8.5~\mu$as and $35.5 \pm 14.9~\mu$as respectively after 15 years.  The velocity magnitude of Sgr A* induced by the $10~\msun$ black holes was on average $0.05 \pm 0.02~\text{km s}^{-1}$, with a perpendicular component of $0.03 \pm 0.01~\text{km s}^{-1}$. For the $100~\msun$ black hole density profile, Sgr A* possessed a mean velocity magnitude of $0.09 \pm 0.04~\text{km s}^{-1}$, and a perpendicular component magnitude of $0.05 \pm 0.02~\text{km s}^{-1}$.

\begin{figure}[h]
	\centering
    \includegraphics[width=0.45\textwidth]{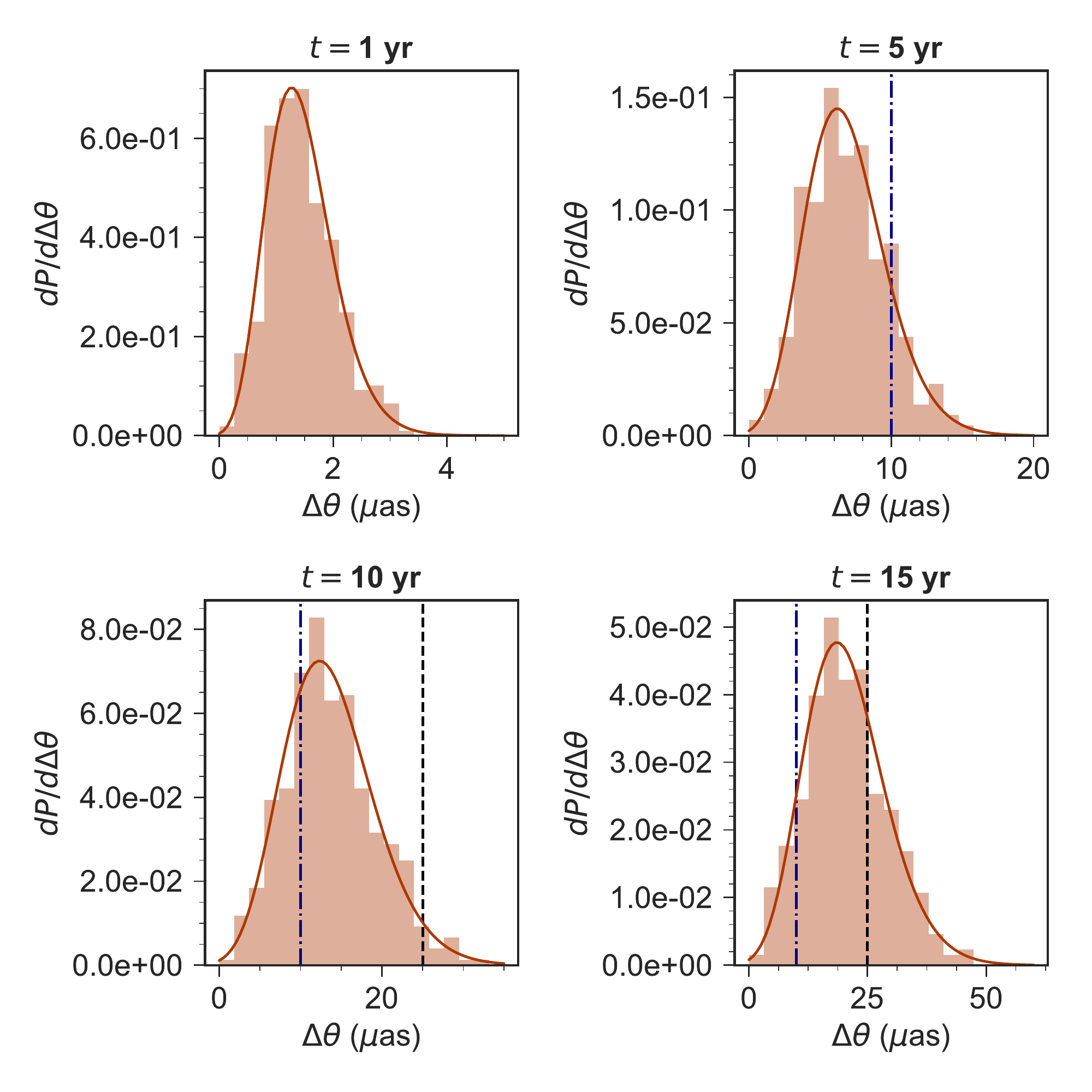}
    \caption{As in Figure \ref{fig:1hist}, with Sgr A* surrounded by a population of $10~\msun$ stellar black holes.}
    \label{fig:10hist}
    
\end{figure}

\begin{figure}[h]
	\centering
    \includegraphics[width=0.45\textwidth]{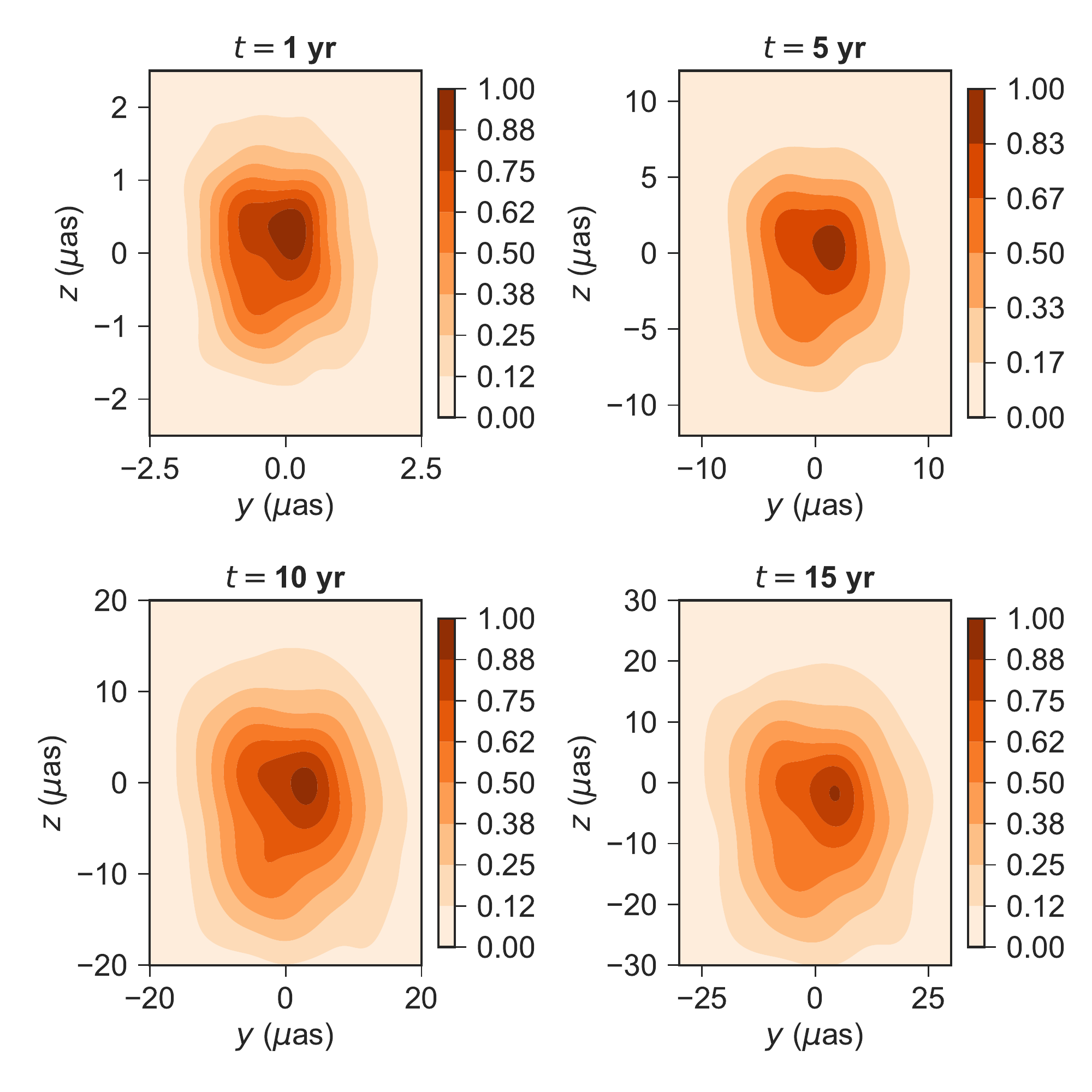}
    \caption{As in Figure \ref{fig:1kde}, with Sgr A* surrounded by a population of $10~\msun$ stellar black holes.}
    \label{fig:10kde}
    
\end{figure}

\begin{figure}[h]
	\centering
    \includegraphics[width=0.45\textwidth]{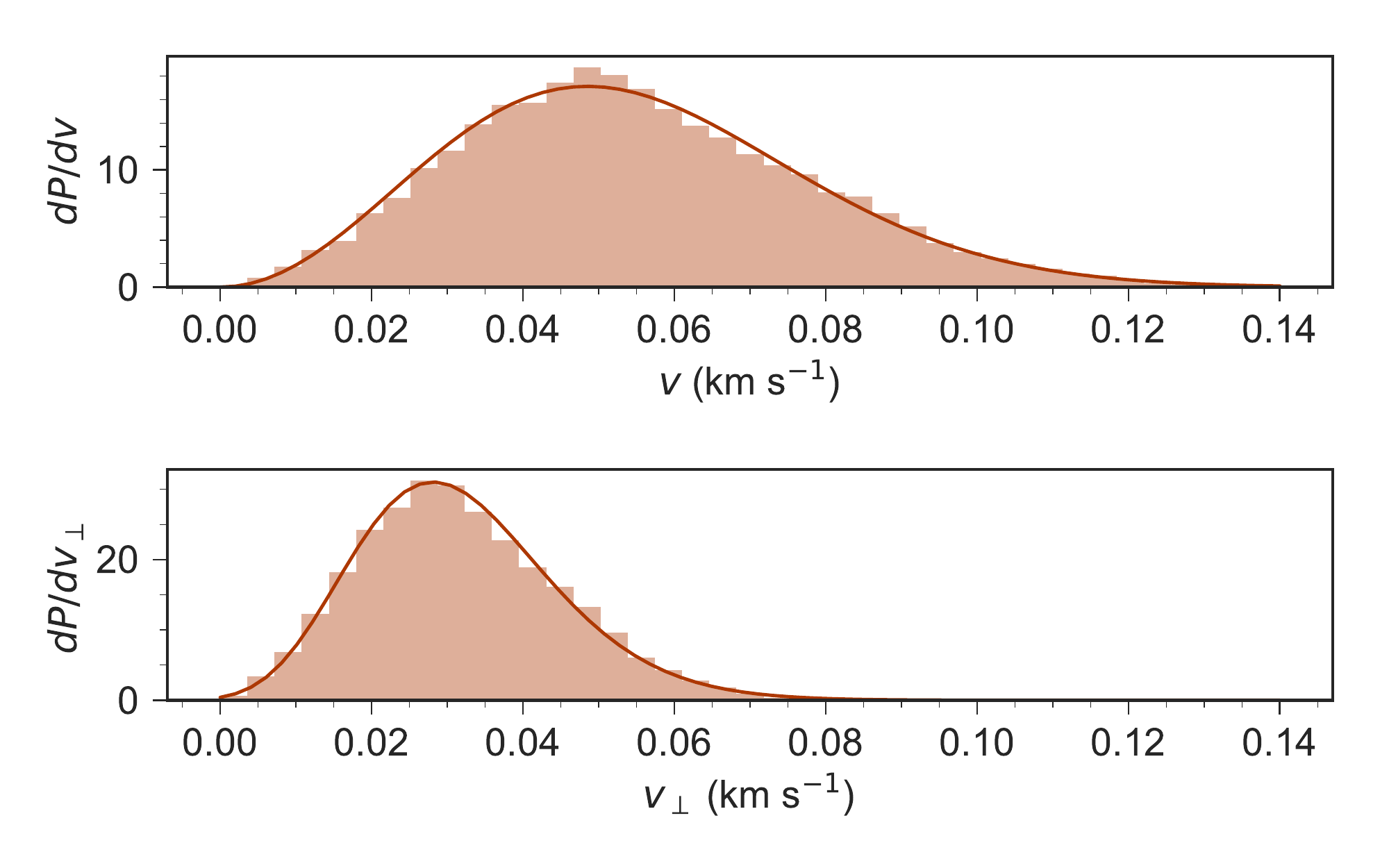}
    \caption{As in Figure \ref{fig:1vel}, with Sgr A* surrounded by a population of $10~\msun$ stellar black holes.}
    \label{fig:10vel}
    
\end{figure}

\begin{figure}[h]
	\centering
    \includegraphics[width=0.45\textwidth]{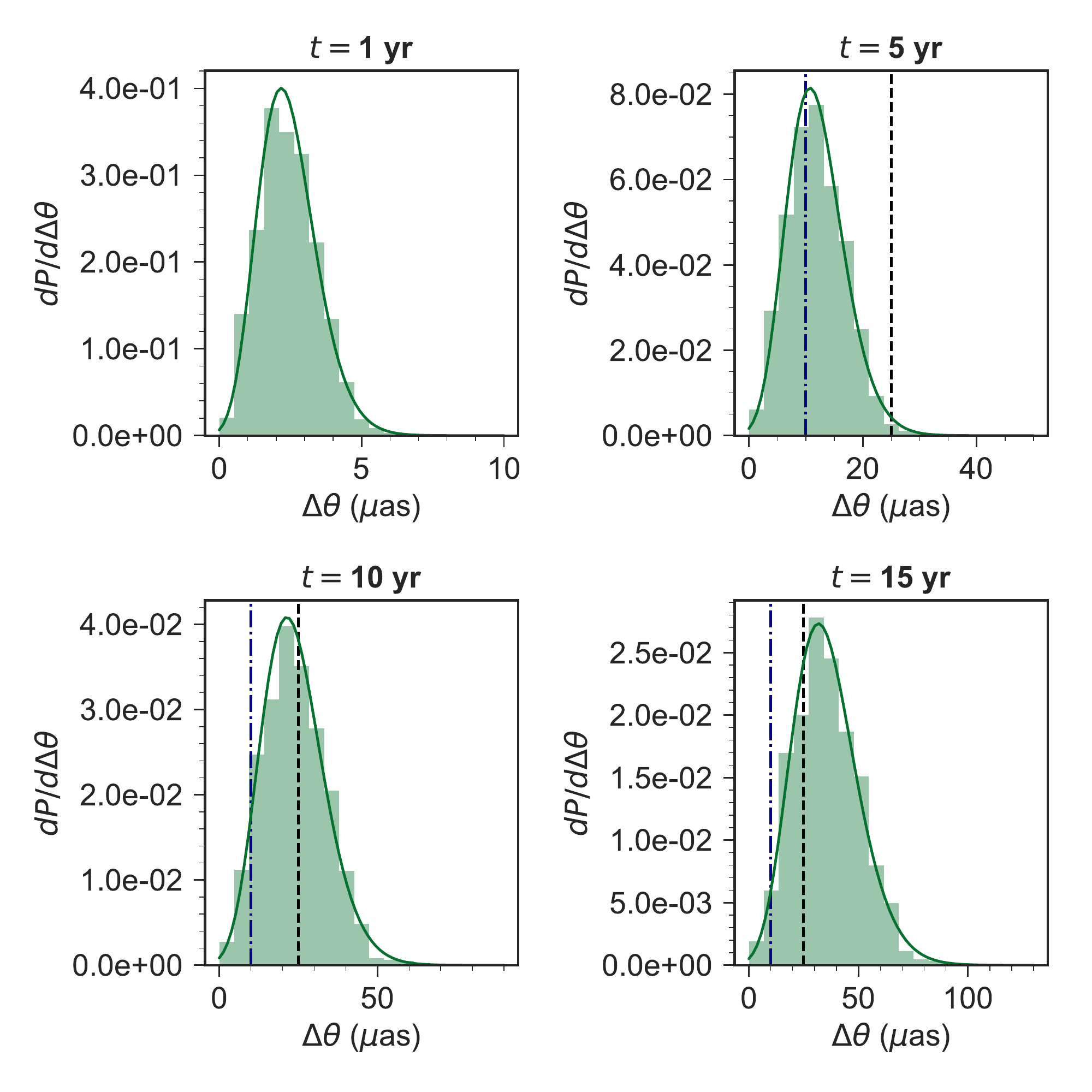}
    \caption{As in Figure \ref{fig:1hist}, with Sgr A* surrounded by a population of $100~\msun$ stellar black holes.}
    \label{fig:100hist}
    
\end{figure}

\begin{figure}[h]
	\centering
    \includegraphics[width=0.45\textwidth]{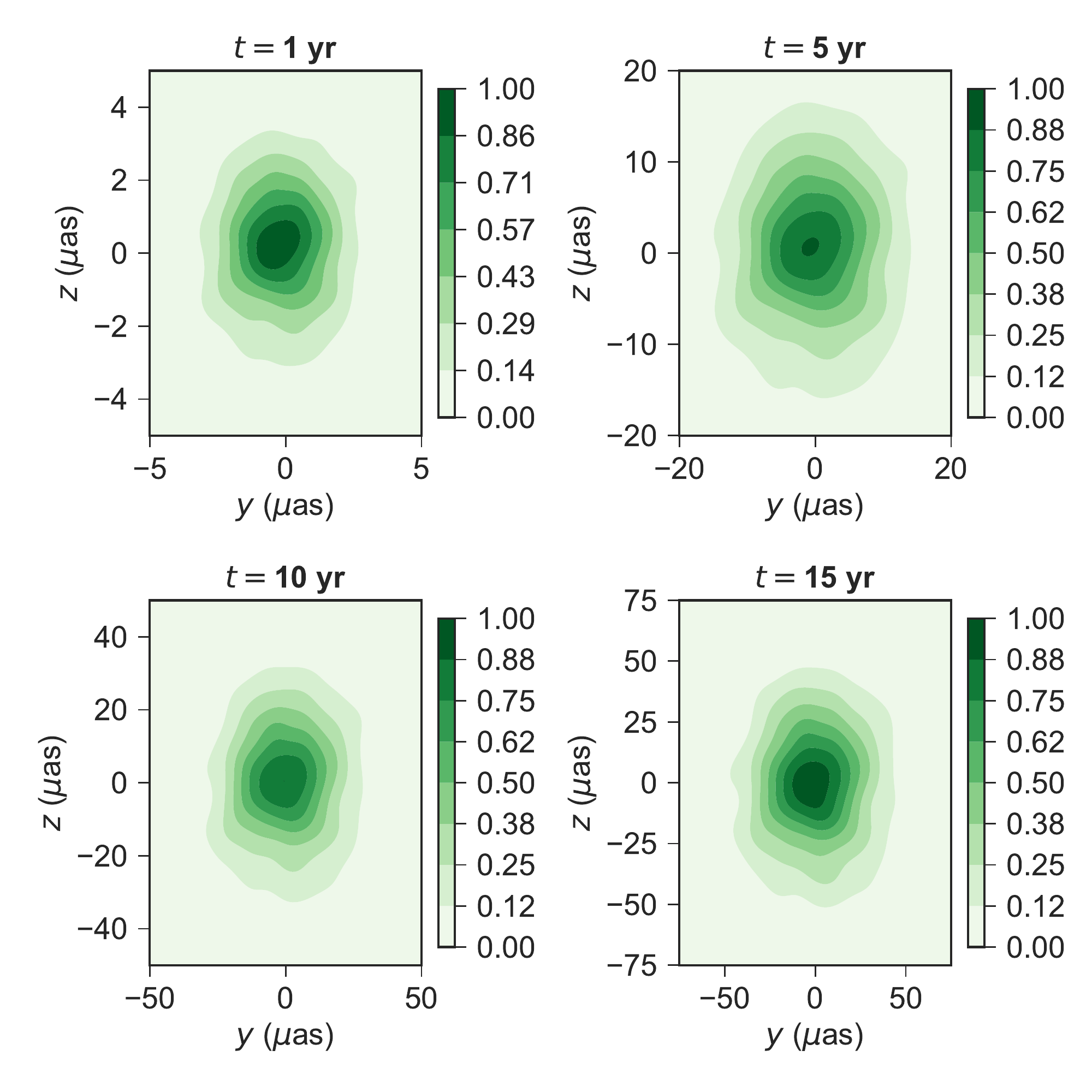}
    \caption{As in Figure \ref{fig:1kde}, with Sgr A* surrounded by a population of $100~\msun$ stellar black holes.}
    \label{fig:100kde}
    
\end{figure}

\begin{figure}[h]
	\centering
    \includegraphics[width=0.45\textwidth]{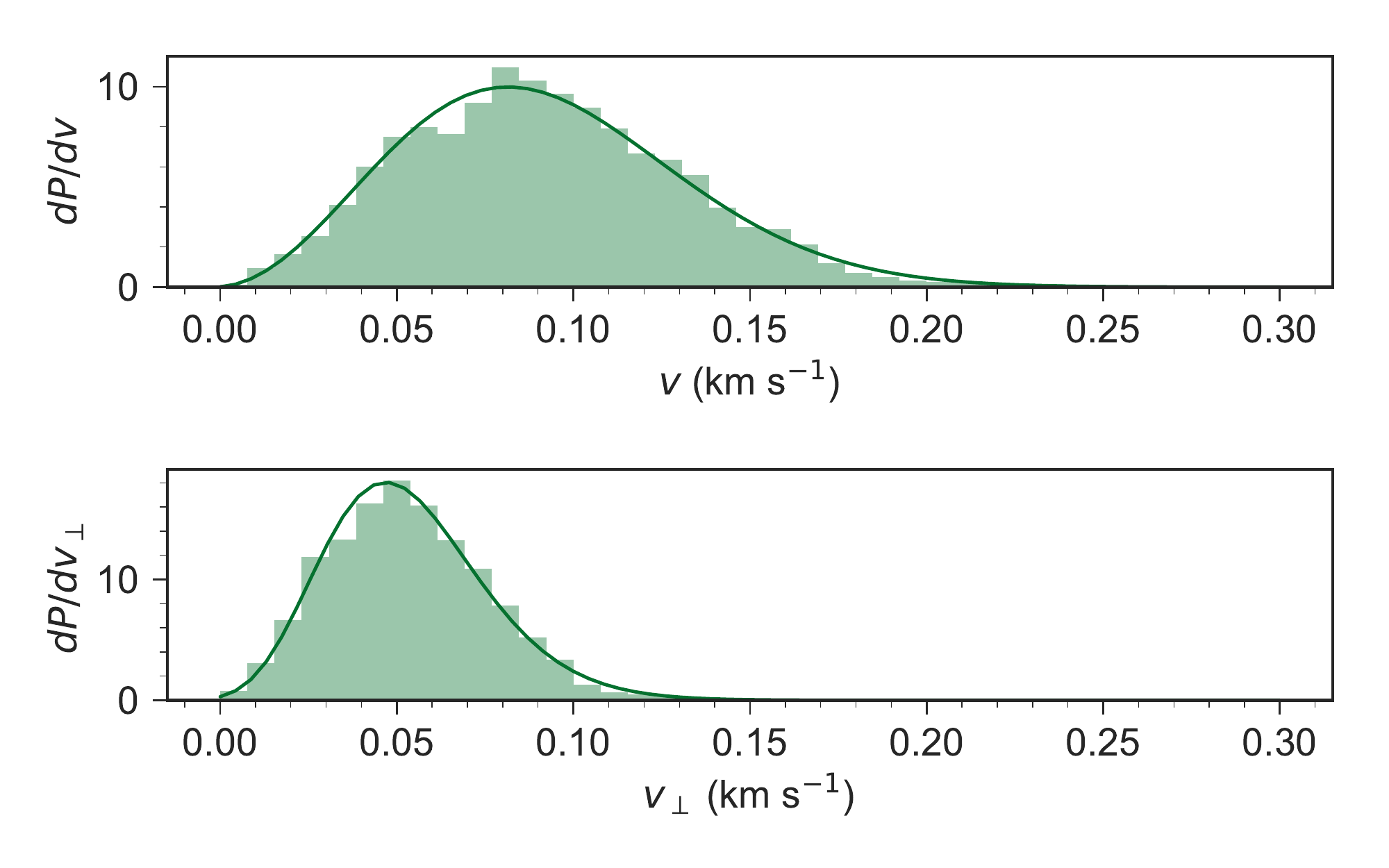}
    \caption{As in Figure \ref{fig:1vel}, with Sgr A* surrounded by a population of $100~\msun$ stellar black holes.}
    \label{fig:100vel}
    
\end{figure}


There were on average $4$ IMBHs of mass $10^3~\msun$ drawn according to our fiducial density profile within one parsec, inducing angular shifts of $39.6 \pm 25.7~\mu$as on the position of Sgr A* after 15  years.  The average velocity of the central black hole induced by this profile was $0.1 \pm 0.06~\text{km s}^{-1}$, though unlike with other density profiles the distribution of this velocity was better fit to a general gamma distribution with mean and standard deviation $0.06\pm 0.04~\text{km s}^{-1}$. The mean perpendicular velocity component was $0.06 \pm 0.04~\text{km s}^{-1}$.

\begin{figure}[h]
	\centering
    \includegraphics[width=0.45\textwidth]{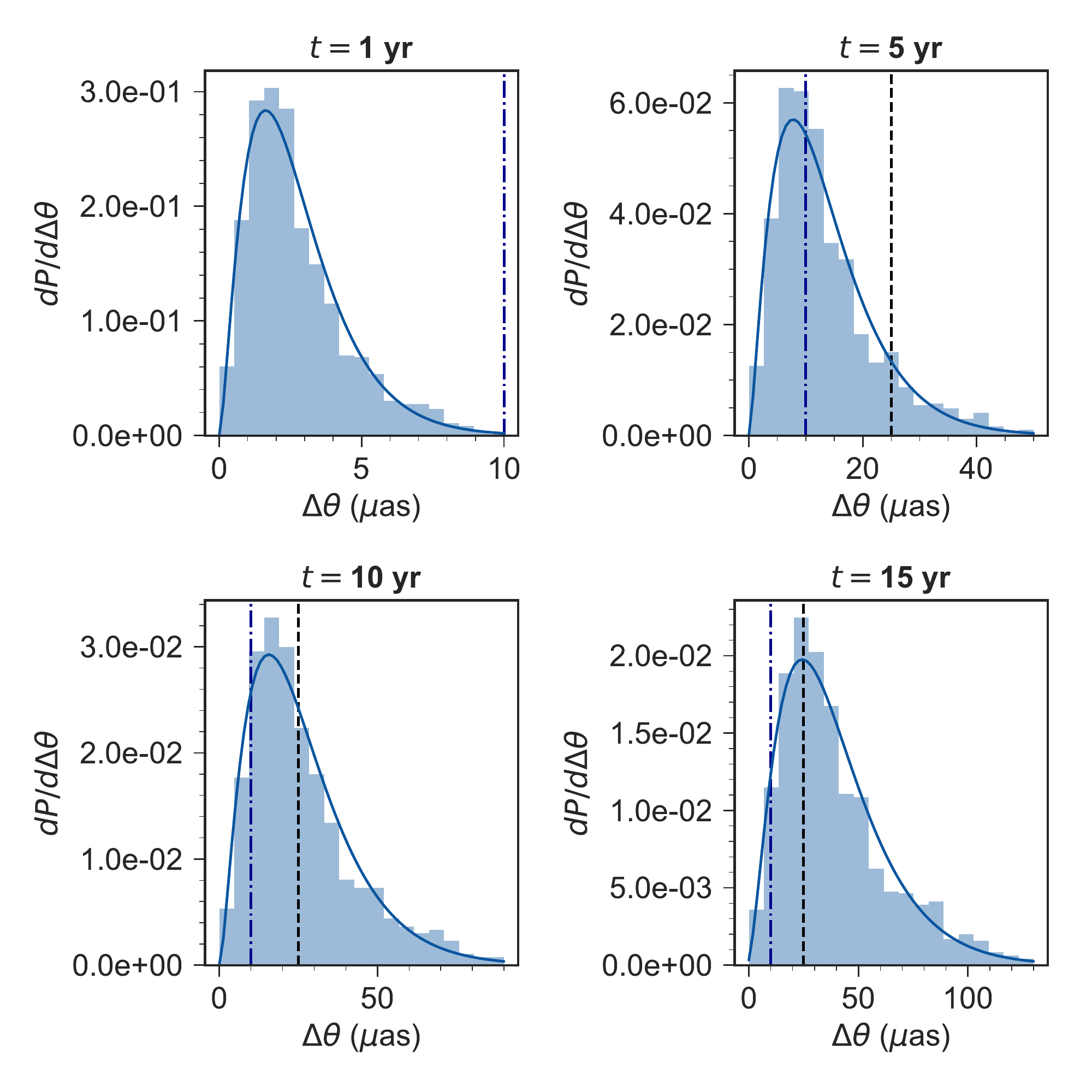}
    \caption{As in Figure \ref{fig:1hist}, with Sgr A* surrounded by a population of $10^3~\msun$ IMBHs.}
    \label{fig:1000hist}
    
\end{figure}

\begin{figure}[h]
	\centering
    \includegraphics[width=0.45\textwidth]{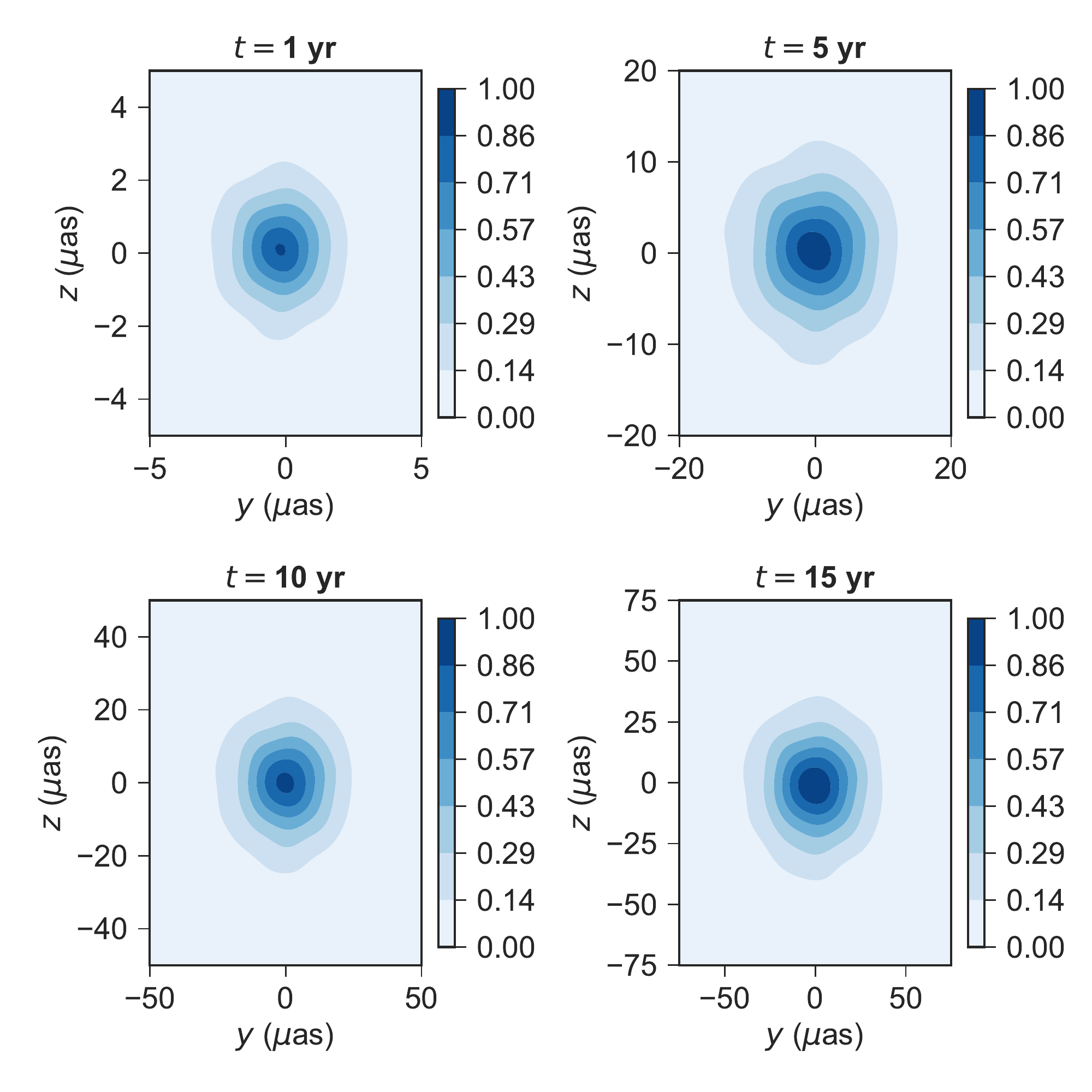}
    \caption{As in Figure \ref{fig:1kde}, with Sgr A* surrounded by a population of $10^3~\msun$ IMBHs.}
    \label{fig:1000kde}
    
\end{figure}

\begin{figure}[h]
	\centering
    \includegraphics[width=0.45\textwidth]{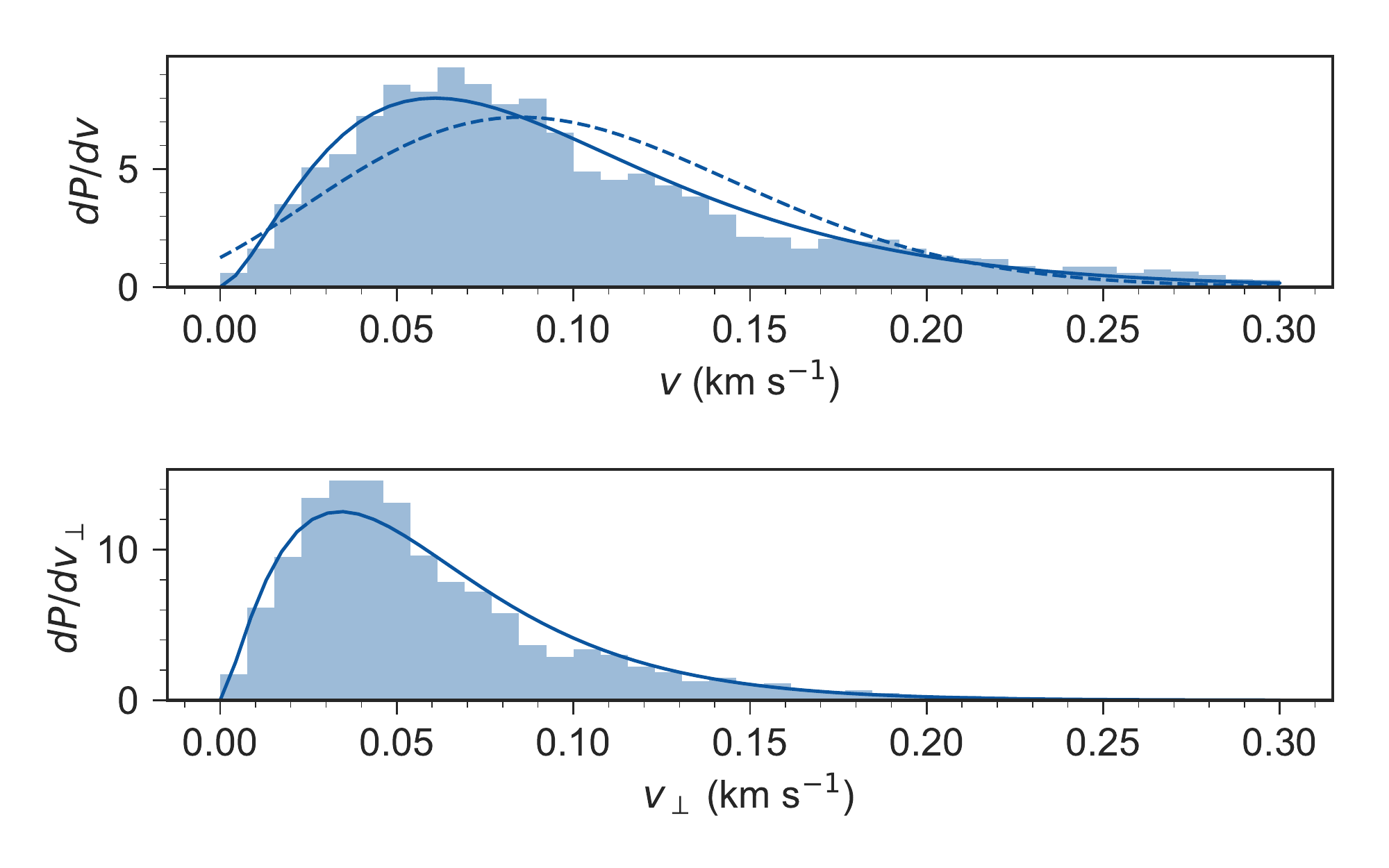}
    \caption{As in Figure \ref{fig:1vel}, with Sgr A* surrounded by a population of $10^3~\msun$ IMBHs. Unlike Figure \ref{fig:1vel}, the solid line in the upper panel outlines a fitted gamma distribution while the dashed line marks a fitted Maxwellian distribution.}
    \label{fig:1000vel}
\end{figure}


From the \citeauthor{Mastrobuono:2014} density profile, an average of ten $10^4~\msun$ IMBHs were drawn in the inner parsec. After 15 years, they induced an angular shift in Sgr A* position of $953.9\pm 459.0~\mu$as, with velocity magnitude and perpendicular component of $2.63\pm 1.33~\text{km s}^{-1}$ and $1.57\pm 0.86~\text{km s}^{-1}$ respectively.

\begin{figure}[h]
	\centering
    \includegraphics[width=0.45\textwidth]{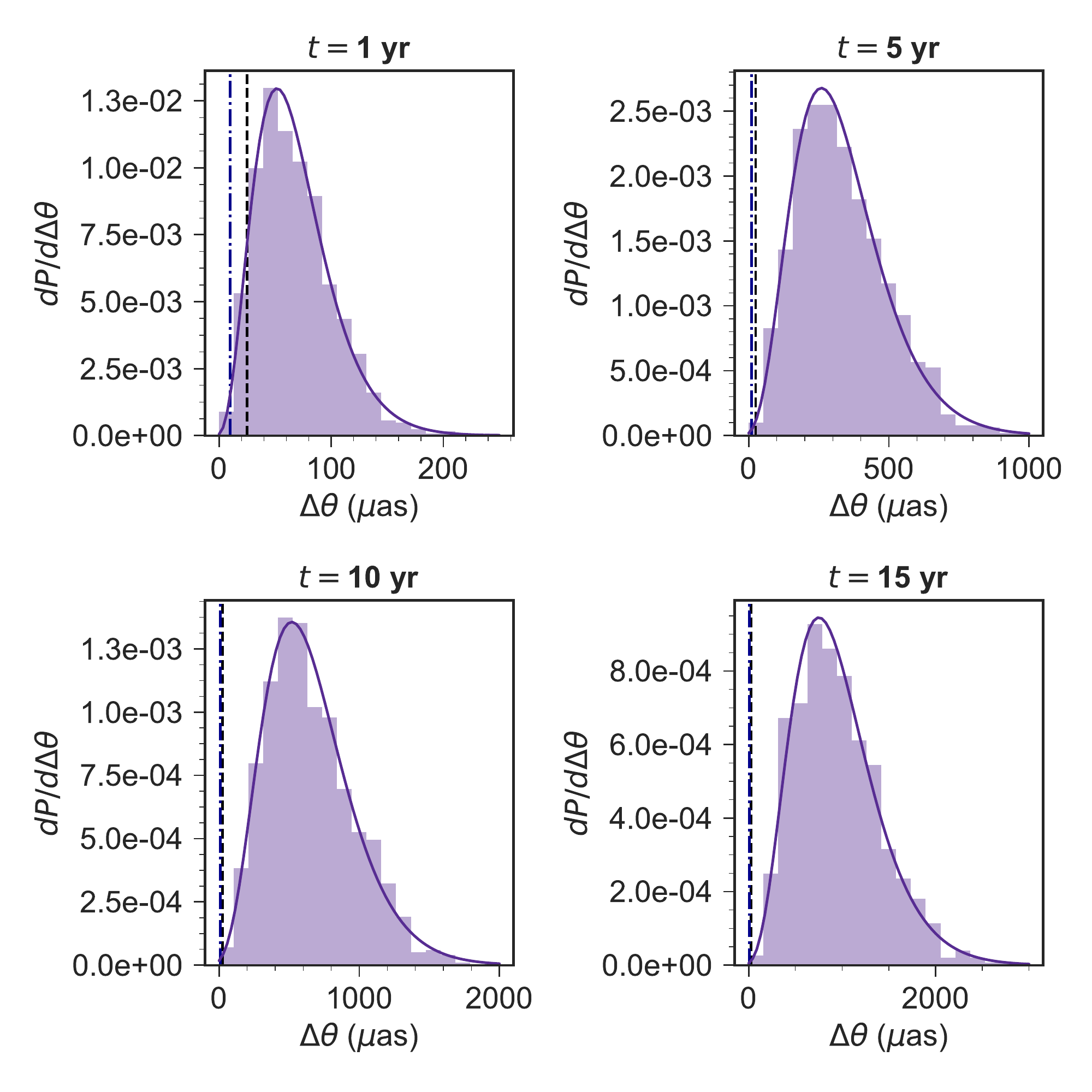}
    \caption{As in Figure \ref{fig:1hist}, with Sgr A* surrounded by a population of $10^4~\msun$ IMBHs.}
    \label{fig:10000hist}
    
\end{figure}

\begin{figure}[h]
	\centering
    \includegraphics[width=0.45\textwidth]{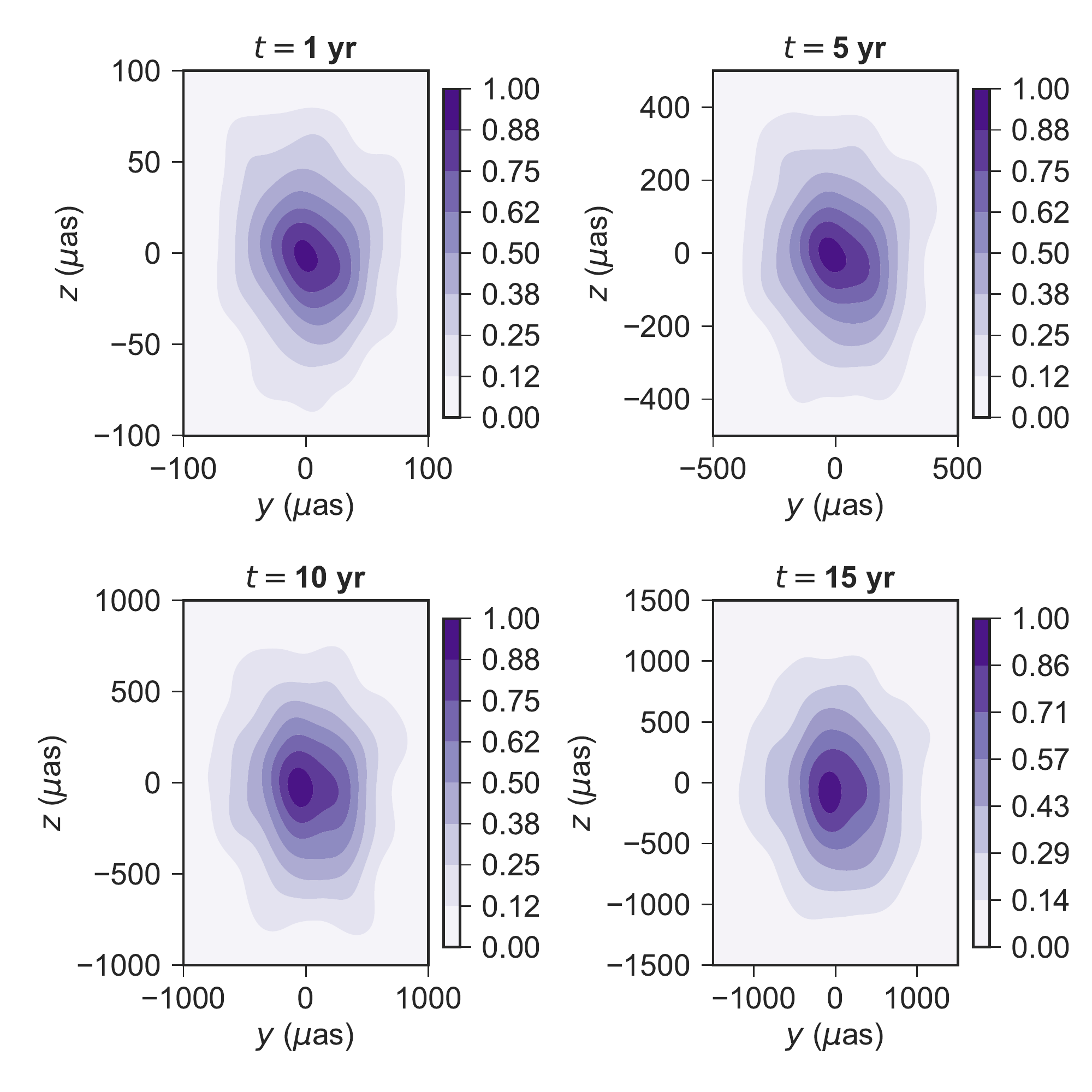}
    \caption{As in Figure \ref{fig:1kde}, with Sgr A* surrounded by a population of $10^4~\msun$ IMBHs.}
    \label{fig:10000kde}
    
\end{figure}

\begin{figure}[h]
	\centering
    \includegraphics[width=0.45\textwidth]{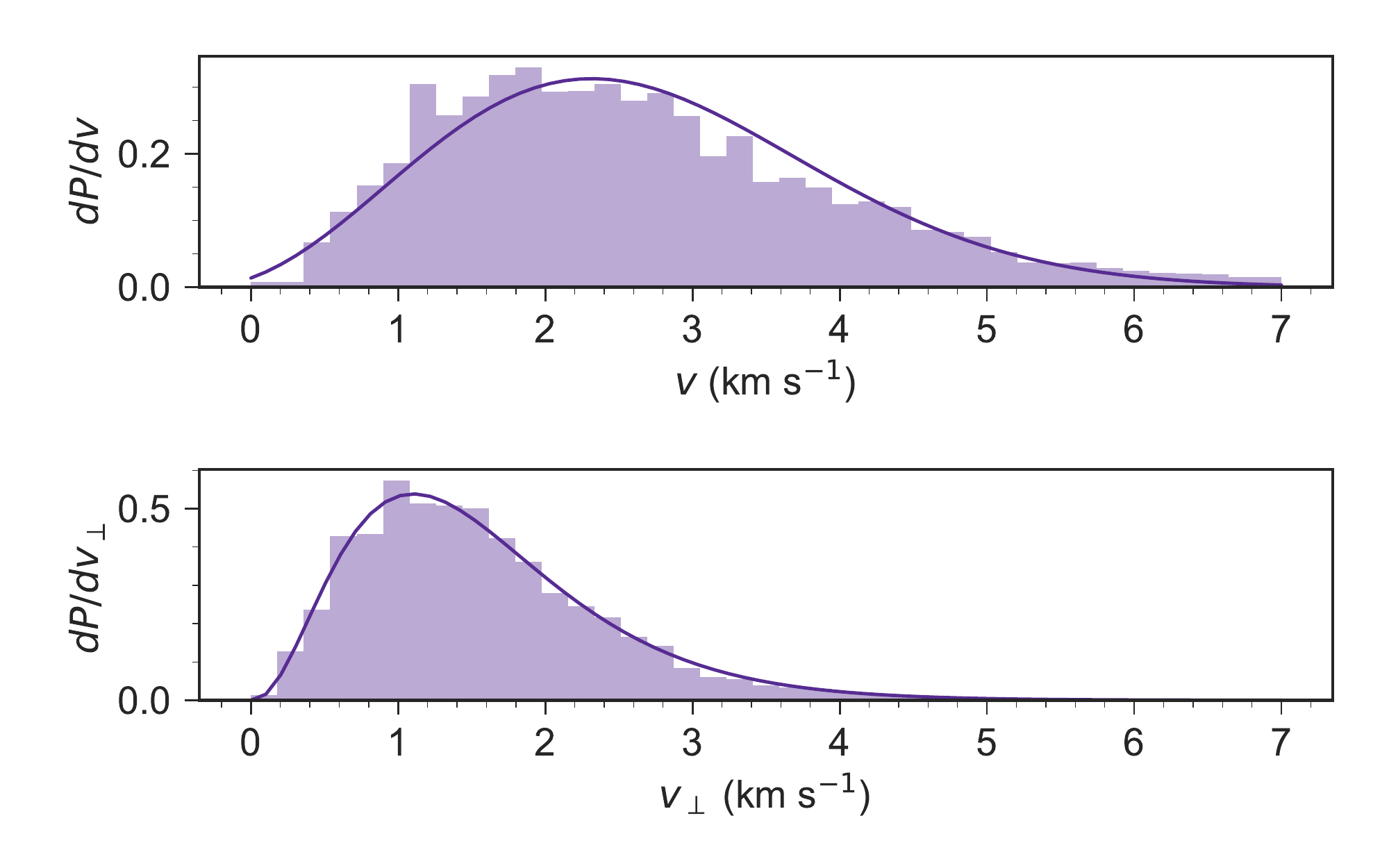}
    \caption{As in Figure \ref{fig:1vel}, with Sgr A* surrounded by a population of $10^4~\msun$ IMBHs.}
    \label{fig:10000vel}
    
\end{figure}

\subsection{Effects on Orbital Parameters of S2}

\begin{figure}
	\centering
    \includegraphics[width=0.50\textwidth]{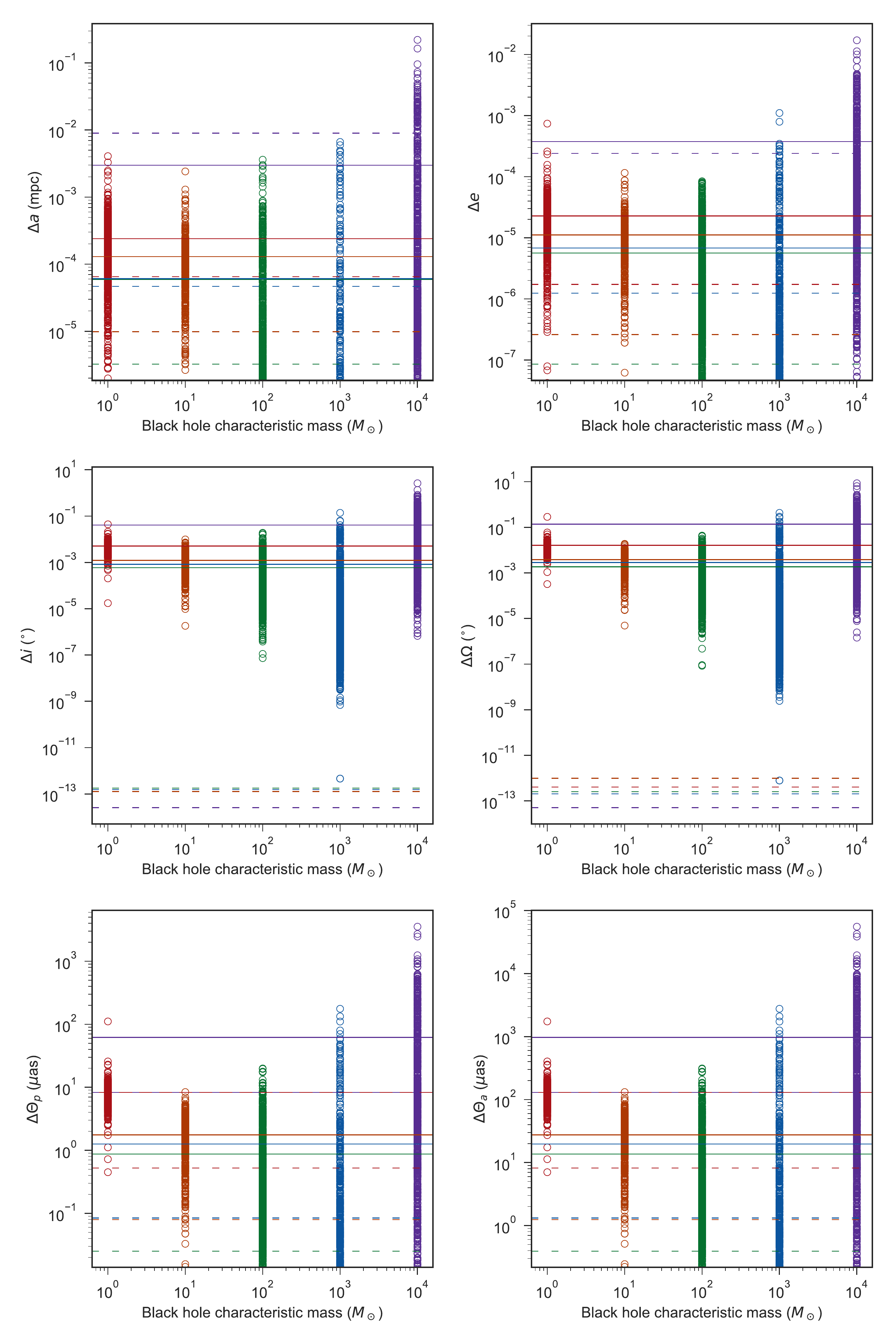}
    \caption[Change in orbital elements of S2 induced by tested density profiles]{Changes in orbital elements (semi-major axis, eccentricity, inclination, longitude of ascending node, shift of periapse, and shift of apoapse) experienced by S2 during each simulation run when surrounded by tested density profiles. Solid lines indicate the average change of the orbital element induced by a certain density profile, with characteristic mass indicated on the x-axis and coordinated by colour.  Dashed lines indicate the average change of the orbital element induced by a smooth version of a certain density profile, also marked by colour.  A single circle indicates the change in orbital element that occurred in one simulation of S2 embedded in a particular density profile.}\label{fig:orbital-elems}
\end{figure}

Table \ref{tab:orbelem} provides a thorough summary of the average changes in semi-major axis ($\Delta a / a$, where $a =4.823~$mpc), eccentricity ($\Delta e$), inclination ($\Delta i$), longitude of ascending node ($\Delta \Omega$), shift of periapse ($\Delta \Theta_p$), and shift of apoapse ($\Delta \Theta_a$) induced by the stellar control and the tested discrete and smooth stellar/IMBH density profiles. For all elements, the mean change did not scale simply with mass; $10^4~\msun$ IMBHs produced the maximal average change in an orbital element, with the stellar control and $10~\msun$ stellar black holes producing the next largest average changes, and the $10^3~\msun$ IMBHs followed by the $100~\msun$ stellar black holes producing the least. Discretely realized profiles induced various mean changes in semi-major axis, differing by approximately a factor of ten ($\Delta a / a \sim 10^{-5} - 10^{-4}$) and possessing no dramatic differences from changes in semi-major axis induced by a smoothly distributed profile.  Stars and low-mass stellar black holes induced changes on eccentricity with a mean of $\sim 10^{-5}$, while for $100~\msun$ and $10^3~\msun$ black holes this value was $\sim 6 \times 10^{-6}$. The mean change in eccentricity due to $10^4~\msun$ black holes was almost a factor of 100 greater at $3.76 \times 10^{-4}$.  The changes in eccentricity induced by discretely distributed particles was less in all cases than those induced by the corresponding smooth simulated profile, though by zero to two orders of magnitude.

\begin{table*}
\centering
\begin{tabular}{@{}llccccc@{}}
\toprule
Orbital element & Mean change & $\langle m_{\rho} \rangle=1~\msun$ & $\langle m_{\rho} \rangle=10~\msun$ & $\langle m_{\rho} \rangle=100~\msun$ & $\langle m_{\rho} \rangle=10^3~\msun$ & $\langle m_{\rho} \rangle=10^4~\msun$ \\ \midrule
\multirow{2}{*}{$\Delta a / a$} & Discrete & 4.14E-05 & 2.07E-05 & 1.22E-05 & 1.26E-05 & 6.21E-04\\
& Smooth & 1.35E-05 & 2.03E-06 & 6.63E-07 & 9.67E-06 & 1.86E-03\\*[5pt]
\multirow{2}{*}{$\Delta e$} & Discrete & 2.28E-05 & 1.12E-05 & 5.61E-06 & 6.81E-06 & 3.76E-04\\
& Smooth & 1.73E-06 & 2.61E-07 & 8.53E-08 & 1.24E-06 & 2.40E-04\\*[5pt]
\multirow{2}{*}{$\Delta i$ (deg)} & Discrete & 5.10E-03 & 1.23E-03 & 5.94E-04 & 8.25E-04 & 4.13E-02\\
& Smooth & 1.27E-13 & 1.27E-13 & 1.78E-13 & 1.53E-13 & 2.54E-14\\*[5pt]
\multirow{2}{*}{$\Delta \Omega$ (deg)} & Discrete & 1.62E-02 & 3.79E-03 & 1.86E-03 & 2.85E-03 & 1.36E-01\\
& Smooth & 4.07E-13 & 9.92E-13 & 2.54E-13 & 2.04E-13 & 5.09E-14\\*[5pt]
\multirow{2}{*}{$\Delta \Theta_p$ ($\mu$as)} & Discrete & 8.3 & 1.8 & 0.9 & 1.3 & 61.9\\
& Smooth & 0.5 & 0.08 & 0.03 & 0.08 & 8.3\\
\multirow{2}{*}{$\Delta \Theta_a$ ($\mu$as)} & Discrete & 130 & 27.6 & 13.7 & 19.8 & 970.4\\
& Smooth & 8.2 & 1.3 & 0.4 & 1.3 & 130.2\\
\bottomrule
\end{tabular}\label{S2-orb-elem2}
\caption{Average changes induced by discrete and smooth density profiles describing the distribution of stars ($\langle m_{\rho} \rangle=1~\msun$), stellar black holes ($\langle m_{\rho} \rangle=10~\msun,\;100~\msun$), and intermediate mass black holes ($\langle m_{\rho} \rangle=1000~\msun,\;10^4~\msun$) on the the semi-major axis ($\Delta a / a$, with $a \sim 5~$mpc), eccentricity ($e$), inclination ($i$), longitude of the periapsis ($\Omega$), angular shift of periapse ($\Delta \Theta_p$), and angular shift of apoapse ($\Delta \Theta_a$) of S2.}
\label{tab:orbelem}
\end{table*}

The second row panels of Figure \ref{fig:orbital-elems} show that smooth profiles had absolutely negligible effects on inclination and longitude of ascending node (all changing these elements by $\sim 10^{-13}$ degrees) in contrast to their discrete counterparts.  Discretely distributed $10^4~\msun$ IMBHs induced the largest average change of $\Delta i = 0.0539^{\circ}$ and $\Delta \Omega = 0.152^{\circ}$. $1000~\msun$ and stellar black holes of masses $10~\msun$ and $100~\msun$ induced average changes of $\Delta i \approx 0.001^{\circ}$ and $\Delta \Omega \approx 0.003^{\circ}$, and the stellar control induced changes in $i$ and $\Omega$ of $\sim 0.003^{\circ}$ and $\sim 0.016^{\circ}$ respectively.

Discretely distributed profiles also generated larger shifts in periapse and apoapse when compared to smooth distributions. For all but the $10^4~\msun$ IMBH profile the mean angular shift of periapse position was $\Delta \Theta_p \lesssim 8~\mu$as.  Shifts in apoapse position $\Delta \Theta_a$ were greater for each profile than $\Delta \Theta_p$, reaching a maximum mean of 0.97 mas with the $10^4~\msun$ IMBHs.  All profiles induced a mean $\Delta \Theta_a$ that was above a $10~\mu$as threshold, reaching a next-largest value of $130~\mu$as with the stellar control, $27.6~\mu$as with $10~\msun$ black holes, $19.8~\mu$as with $1000~\msun$ black holes, and $13.7~\mu$as with $100~\msun$ black holes.

\section{Discussion}
\label{sec:5}

Though the simulations carried out in this work were based on simplified assumptions, their results provide non-trivial insight into which black hole populations may be detectable in the Galactic Centre.  Below we compare obtained results to our strongest current observational constraints, which rely on the proper motion of Sgr A* possessing a $2 \sigma$ upper limit of $1.8~\text{km s}^{-1}$ for the velocity perpendicular to the galactic plane \citep{Reid:2004}, a $\sigma$ upper limit of $1.3~\text{km s}^{-1}$ in the direction of galactic rotation \citep{Reid:2009}, and the constrained error of \citet{Gillessen:2009} orbital fit for S2.

With the constraint of proper motion in mind, all stellar mass and intermediate mass black hole profiles under consideration are allowed. The combined precision of the EHT in sub-array mode \citep[$\gtrsim 25~\mu$as, see][]{Broderick:2011} would allow detection within 15 years of the angular displacement induced on Sgr A* by any of the examined profiles.  The velocity kick given to on Sgr A* by stars and stellar black holes ($\sim 0.06~\text{km s}^{-1}$) stand in agreement with an amplitude of Brownian motion calculated by \citet{Merritt:2007, Chatterjee:2002, Loeb:2013}, using $\langle m_* \rangle = 1~\msun$ and $\mbh = 4.4 \times 10^6~\msun$ (See Equation (1.18)), as well as $N$-body simulations of stars within the inner $2$ pc performed by \citet{Reid:2004}.  Both $10^2~\msun$ and $10^3~\msun$ black holes induce angular shifts in the position of Sgr A* at a rate of $\gtrsim 2~\mu\text{as yr}^{-1}$.  The mean velocity of Sgr A* when surrounded by each profile is on the order of $0.1~\text{km s}^{-1}$, closer to estimates on velocity out of the disk as outlined by \citet{Reid:2004} than results from other profiles.  However, due to their similar dynamical signature the tested profiles for $10^2~\msun$ and $100^2~\msun$ black holes are arguably impossible to differentiate from their gravitational effect on Sgr A* alone.  The $10^4~\msun$ IMBH profile produced the largest angular shifts, with a mean of $\sim 65~\mu\text{as yr}^{-1}$.  It additionally induced the largest intrinsic velocity of $2.62 \pm 1.35~\text{km s}^{-1}$.  The perpendicular component of this was $1.6 \pm 0.9~\text{km s}^{-1}$, the mean of which is just within the present $2\sigma$ limit of $1.8~\text{km s}^{-1}$ \citep{Reid:2004}.

To properly examine the impact of these density profiles on S2, we first review the current accuracies in VLT and Keck data embedded in errors allotted for by \citet{Gillessen:2009}.  The best fitted values for semi-major axis and eccentricity are accurate to $\sim 2~$mas and $3 \times 10^{-3}$ respectively, and the inclination, longitude of ascending note, and argument of periapse are accurate to $0.72^{\circ} - 0.81^{\circ}$.  To properly use the orbit of S2 as a probe for stellar remnants and IMBHs, the changes induced by these dark objects must necessarily fit within these accuracies while additionally surpassing orbital changes associated with relativistic effects.  \citet{Grould:2017} have investigated the potential of GRAVITY to detect various relativistic effects on the S2, concluding that a shift of periapse due to relativistic advance will be $30~\mu\text{as}$ in $14~$years (roughly $34~\mu$as per revolution).  Alternatively \citet{Gualandris:2010} calculated the displacement in the star's apoapse as
\begin{equation}
  \Delta r_a \approx a(1+e)\Delta \varpi \approx \frac{6\pi G \msgr}{c^2(1-e)},
\end{equation}
where $\Delta \varpi = \Delta (\Omega + \omega)$ is the advance in periapse angle. Using $M_{\rm Sgr} = 4.4 \times 10^6~\msun$ and $R_0 = 8~$kpc, $\Delta r_a$ subtends an angle on the sky, $\Delta \Theta_a = 0.86~$mas. The angle on the sky subtended by the displacement in the star's periapse can then be calculated as $\Delta \Theta_p = \Delta \Theta_a(1-e)/(1+e) \approx 54.9~\mu$as. Thus in $\Delta \Theta_a$ and $\Delta \Theta_p$, along with constrained errors of $a$, $e$, $i$, $\Omega$, and $\omega$, we have lower limits and upper limits respectively to determine the detectability of a certain density profile's dynamical signature on S2.

There is a notable difference between the dynamical effect of smooth and discrete distributions in orbital arguments pertaining to angle ($i$ and $\Omega$). A smooth or centralized distribution creates virtually no changes in these parameters, whereas the gravitational effect of discretely distributed particles generates distinct apsidal precession of the orbit of S2.  The average change in $i$ and $\Omega$ for all density profiles are within current error of $0.7^{\circ}$.  It is interesting to note that the mean change in inclination and longitude of ascending node due to the stellar control and $10~\msun$ stellar mass black holes exceeds that induced by $10^2~\msun$ stellar black holes and $10^3~\msun$ IMBHs.  This may be because, due to their increased number density, these smaller objects experience closer encounters with S2 that more intensely perturb its orbit.  However, stars, stellar black holes, and $10^3~\msun$ IMBHs all induced changes in periapse and apoapse under estimated relativistic shifts, implying that these density profiles are not detectable through observation of S2's orbital precession.  Only the $10^4~\msun$ IMBH profile induced changes in these parameters exceeding the lower limit placed by the relativistic effects. With this, S2 appears to be a more effective probe for detecting IMBHs of mass $\gtrsim 10^4~\msun$.  Particularly through observations of $\sim$ milli-arcsecond angular shifts in apoapse, along with perceptible changes in inclination and longitude of ascending node, we can infer the existence of surrounding discrete objects.

This work can be expanded upon by an inclusion of a larger variety of density profiles and improved numerical methods. An asymmetric distribution of particles would generate a larger amplitude wobble; this may have already played a role in simulations of the $10^4~\msun$ IMBH profile, where the assumed spherical symmetry was compromised by a relatively small number density and larger gravitational effects on Sgr A* and S2 were generated. Even though this work identified a potentially observable IMBH black hole density profile through its gravitational effects on S2, this effect may be replicable by alternative unseen matter or a single large IMBH.  Thus, it is critical to understand how the magnitude of these changes evolve over multiple periods of S2's orbit.


\section*{Acknowledgements}


This work was supported in part by the Mellon Mays Foundation and by the Black Hole Initiative at Harvard University, which is funded by a grant from the John Templeton Foundation.




\bibliographystyle{mnras}
\bibliography{references} 






\bsp	
\label{lastpage}
\end{document}